\documentclass[%
 reprint,
 twocolumn,
 amsmath,amssymb,
 aps,
floatfix
]{revtex4-2}

\usepackage{graphicx}
\usepackage{dcolumn}
\usepackage{bm}
\usepackage{algorithm}
\usepackage{algpseudocode}
\usepackage{textcomp}

\begin{document}

\preprint{APS/123-QED}

\title{Hardware-efficient random circuits to classify noise in a multi-qubit system}

\author{Jin-Sung Kim}
\email{jin@ibm.com}
 \affiliation{IBM Quantum}
 
\author{Lev S. Bishop}%
 \affiliation{IBM Quantum}
\author{Antonio D. C\'{o}rcoles}
 \affiliation{IBM Quantum}
\author{Seth Merkel}
 \affiliation{IBM Quantum}
\author{John A. Smolin}
 \affiliation{IBM Quantum}
\author{Sarah Sheldon}
 \affiliation{IBM Quantum}

\date{\today}

\begin{abstract}
In this work we extend a multi-qubit benchmarking technique known as the Binned Output Generation (BOG) in order to discriminate between coherent and incoherent noise sources in the multi-qubit regime. While methods exist to discriminate coherent from incoherent noise at the single and few-qubit level, these methods scale poorly beyond a few qubits or must make assumptions about the form of the noise. On the other end of the spectrum, system-level benchmarking techniques exist, but fail to discriminate between coherent and incoherent noise sources. We experimentally verify the BOG against Randomized Benchmarking (RB) (the industry standard benchmarking technique) in the two-qubit regime, then apply this technique to a six qubit linear chain, a regime currently inaccessible to RB. In this experiment we inject an instantaneous coherent $Z$-type noise on each qubit and demonstrate that the measured coherent noise scales correctly with the magnitude of the injected noise, while the measured incoherent noise remains unchanged as expected. This demonstrates a robust technique to measure coherent errors in a variety of hardware.

\end{abstract}

\maketitle

A fundamental barrier in achieving quantum advantage in quantum computers is the presence of noise, which reduces the length and complexity of computations achievable by current quantum processors \cite{jurcevic2020,cross2019_qv}. Noise can be broadly categorized as incoherent noise, e.g. stochastic noise originating in relaxation ($T_1)$ and dephasing ($T_2$) events \cite{catelani2012_quasiparticle,krantz2019_quantum_engineers}, or coherent (unitary) noise, e.g. repeatable over/under rotations caused by miscalibrations or cross-talk \cite{sheldon2016_iterative,sheldon2016_hamtomo}. In superconducting qubit systems, incoherent noise is often attributed to microscopic materials defects \cite{lisenfeld2016, place2020}, quasi-particles \cite{deGraafe2020,catelani2012_quasiparticle, barends2011}, and coupling to environmental noise \cite{rigetti2012,barends2011} whereas coherent noise is often due to unwanted qubit-qubit interactions and microwave cross-talk from control lines \cite{malekakhlagh2020,sheldon2016_hamtomo}. As such, the level of incoherent noise present in a device can generally not be improved once the device is fabricated and cooled down but coherent noise can often be mitigated by control techniques \cite{ball2020software, goerz2017}, echo pulses \cite{wei2020_ghz,sundaresan2020,kandala2020_cnot}, and noise tailoring  \cite{hashim2020,wallman2016}. 

Importantly, the resulting algorithmic error scales differently depending on the type of noise present in the system. Coherent noise results in a worst-case quadratic error scaling in the circuit size (both in the length of the circuit and number of qubits), whereas incoherent noise results in a linear error increase \cite{sheldon2016_iterative, wallman2016,Iverson_2020}. Therefore, for near-term applications, the presence of coherent noise within a given system is more detrimental to the algorithmic fidelity, at least before fault-tolerant quantum computing is achieved \cite{Iverson_2020}. Once coherence times allowing for fault-tolerant quantum computing are achieved, though, it will be paramount to ensure coherent errors are minimal and uncorrelated enough to maintain gate error rates below the fault-tolerance threshold. 

While a mature suite of hardware benchmarks exists to characterize qubits and gates, these methods tend to scale poorly beyond roughly three qubits \cite{McKay19}, at least at current coherence and control limits. For example, variants of tomography can be used to fully characterize the quantum process and state but are susceptible to state preparation and measurement errors (SPAM) and the number of experiments required to characterize the system scales exponentially with the number of qubits. Randomized benchmarking~\cite{Magesan11} is currently the industry standard for measuring single and two-qubit gate errors and is robust to SPAM errors.
While standard RB gives only a single noise parameter, it can be modified to detect coherent errors~\cite{Magesan12,sheldon2016_iterative}, characterize the error of a specific gate~\cite{Magesan12}, as well as measure leakage out of the computational space~\cite{Andrews19}. However, implementing RB beyond a few qubits is difficult because the depth of a $n$-qubit Clifford is a quadratic function of $n$ which limits the practical application of multi-qubit Cliffords in current technologies, and with typical gate errors, to at most three or four qubits \cite{McKay19}.  Characterizations on larger numbers of qubits are of immediate practical interest in order to gauge the efficacy of near-term quantum computations. Additional noise sources may also be revealed in multi-qubit scenarios that may go undetected simply looking at the isolated one- and two-qubit error rates present in the system~\cite{McKay19,mckay2020correlated}.
    
    
    
For system-level characterization, the Quantum Volume (QV) \cite{jurcevic2020, cross2019_qv} remains the gold standard for hardware-agnostic performance benchmarking and incrementing QV signifies meaningful system improvements. We can ask, however, within a single step of QV, can we implement less demanding random circuits to extract fine-grained information which discriminates between coherent and incoherent noise? Here we introduce ``hardware-efficient random circuits,'' a family of circuits that exploits the native interactions available in the hardware, and implement a binning method introduced by Bouland \textit{et al.} called ``Binned Output Generation'' (BOG)~\cite{Bouland19} which allows us to characterize the output of these circuits. We extend this method and show that by binning the same experimental results in two separate ways we can discriminate between coherent noise and incoherent noise in a multi-qubit system, thus extending noise classification beyond just the few qubits practically available hitherto.

In the following, we describe the hardware-efficient random circuits we implement as well as the binning process. We experimentally verify this technique in the two-qubit regime and compare the CNOT rate extracted from the BOG to the CNOT rate from simultaneous two-qubit RB. We then perform a six-qubit BOG experiment and extract an average CNOT error rate which is slightly higher than the average CNOT error rate extracted from the two-qubit experiment, indicating additional sources of error not detected by the two-qubit measurement. In this regime we re-bin the experimental results in order to extract only the incoherent error. The average incoherent error per CNOT agrees well with measurements of the purity error measured by purity RB \cite{wallman2015}. Purity RB is a method of measuring how coherent the state is by applying state tomography to the end of a RB sequence and measuring the magnitude of the Bloch vector. Finally, we inject a purely coherent $Z$ noise source \cite{mckay2017_zgates} into the six-qubit circuits and extract average CNOT error rates consistent with the injected $Z$ noise with an additional static $ZZ$ component of the same order as the independently-measured $ZZ$ interaction. Importantly, the measured incoherent noise remains constant during this experiment, demonstrating that our re-binning technique only detects the incoherent error and is robust to changes in coherent error.

Hardware-efficient random circuits are a family of circuits which respect the native connectivity and native entangling gate of the hardware. They consist of alternating layers of Haar-random single qubit rotations and two-qubit entangling gates between coupled qubits. For our architecture, the CNOT is the native entangling gate. Entangling gates are applied to alternating adjacent pairs of qubits, i.e. on one cycle qubit $i$ would be entangled with its neighbor $i+1$, and on the next cycle with $i-1$. In this way, these circuits are agnostic to the details of the hardware architecture and thus can be applied to a variety of different systems.  

In a perfect, noiseless system, with sufficiently many cycles of the aforementioned operations, the resulting quantum state will land at a random point within the Hilbert space. Projecting this state onto a measurement basis results in series of bitstrings whose probability distribution will tend towards a Porter-Thomas distribution ~\cite{PT56, Neill18}. For Hilbert spaces with dimension much larger than order unity, the Porter-Thomas distribution is well approximated by an exponential decay function.

The premise of the BOG, and other similar metrics like the cross-entropy~\cite{Neill18} and the heavy output generation (HOG) \cite{jurcevic2020,cross2019_qv}, is to pre-compute (with a classical computer) the ideal (i.e. noiseless) probability distribution that results from executing the circuit and compare this ideal distribution to the noisy distribution produced by the quantum hardware. The main benefit these methods provide is that these quantities can be estimated efficiently, in terms of quantum resources, with a non-exponential number of experiments ~\cite{Bouland19,Neill18}, though the classical pre-computation still requires exponential resources. 


We compute a fidelity score for the BOG based on how far the noisy distribution is from the ideal (noiseless) distribution and normalize said fidelity taking into account the other limiting case of an incoherent mixture. This is fidelity is computed in the following way. Given a target circuit described by the unitary matrix $\hat{U}$, we define the outcomes for the ideal probability distribution as 

\begin{equation}
p_i = |\langle x_i \vert \hat{U} \vert 0 \rangle|^2,
\end{equation}
where $x_i \in {\mathbb Z}^n$ denote the possible output bitstrings from the circuit for $n$ qubits. In the presence of noise, the experimentally measured probabilities $q_i$ will deviate from the ideal probabilities.

To estimate the noisy probability distribution using the BOG, the probability space spanned by $[0,1]$ is divided into poly($n$) number of bins \cite{Bouland19}. The distribution is reconstructed by adding each experimentally measured probability $q_i$ to the bin which contains the bitstring $x_i$'s ideal probability $p_i$ (algorithm \ref{bog_algo_sv}). For each bin $[a/N,b/N]$, where $N=2^n$, the bin edges are constructed such that $\int_{a}^{b}q e^{-Nq}= \Theta (1/{\rm poly}(n))$ \cite{Bouland19}. This ensures that for an ideal Porter-Thomas distribution, the weights in each bin are equal. It is worth noting that the cross-entropy and HOG can be viewed as two limiting cases of the BOG \cite{Bouland19}. For the HOG, outcomes are sorted into just two bins (heavy or not heavy) and for the cross-entropy, each outcome is placed into its own bin. The BOG maximizes the amount of information obtained from the pre-computed distribution and can rule out ``imposter'' distributions \cite{Bouland19}.


To give an example of the binning procedure, suppose the all 0 bitstring's ($x_0$) ideal probability ($p_0$) is 0.05 which happens to be contained in the first bin ($bin_0$), spanning $[0,0.1]$. The experimentally measured probability $q_0$ of $x_0$ is then added to $bin_0$. This process is repeated for each bitstring. In the presence of noise, the measured probability for each bitstring will deviate from the ideal, a distribution that approaches the uniform distribution, and the weights of each bin will deviate away from the ideal weight.

\begin{algorithm}[H]
  \caption{Binned Output Generation}
  \label{bog_algo_sv}
   \begin{algorithmic}[1]
   \For{\textbf{$x_i$} in \textbf{$x_{all}$}}
        \State $p_i$ = ideal\_probabilities [$x_i$]
        \State $q_i$ = experimental\_probabilities [$x_i$]
        \State bin\_index = assign\_bin($p_i$)
        \State bins [bin\_index] += $q_i$
    \EndFor
   \end{algorithmic}
\end{algorithm}

\begin{algorithm}[H]
  \caption{Binned Output Generation for Incoherent Errors}
  \label{bog_algo_exp}
  \begin{algorithmic}[1]
  \For{\textbf{$x_i$} in \textbf{$x_{all}$}}
        \State $q_i$ = experimental\_probabilities [$x_i$]
        \State bin\_index = assign\_bin($q_i$)
        \State bins [bin\_index] += $q_i$
    \EndFor
  \end{algorithmic}
\end{algorithm}

To convert the measured frequencies to a fidelity we use the following expression 
\begin{equation}
        \text{Fidelity}=\left\langle1-\frac{\left|{\bf bins}(\psi_{\rm ideal})-{\bf bins}(\rho_{\rm exp})\right|_1}{\left|{\bf bins}(\psi_{\rm ideal})-{\bf bins}( {\mathbb I}/2^n )\right|_1}\right\rangle_{\hat{U}},
        \label{eq:fidelity}
\end{equation}

\begin{figure*}[t!]
    \centering
    \includegraphics[width=0.9\textwidth]{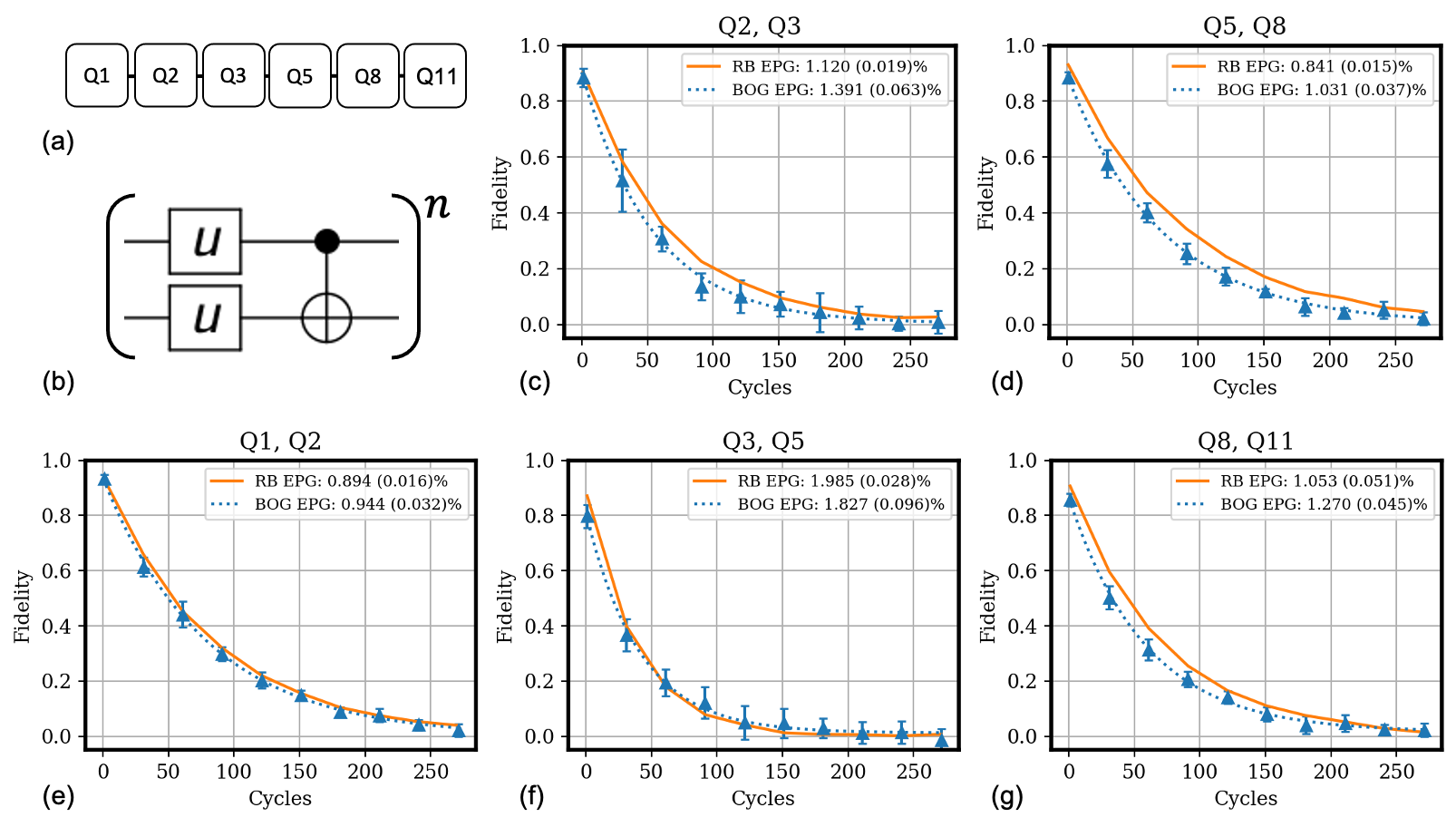}
    \caption{(a) Connectivity diagram of the qubits used in this work. (b) One cycle of the two-qubit BOG circuit. Each $u$ gate denotes a unique Haar-random single-qubit rotation, randomized for each gate for a given circuit seed. (c)-(g) Plots of the two-qubit BOG fidelity as a function of the circuit depth. The blue dotted lines denote exponential fits to the data. Error bars are computed by bootstrapping 10 groups of nine seeds together and computing the standard deviation of these values. Orange lines denote noisy simulations of the BOG circuits where the only noise parameters included are the readout error and the CNOT error measured by RB. Single-qubit errors are omitted from the simulation as the single-qubit error rate is typically one to two orders of magnitude lower than the CNOT error.}
    \label{fig:2q_bog}
\end{figure*}

that is the 1-norm distance of the binning of the experimental data ($\rho_{\rm exp}$) from the binning of the ideal output, normalized by the difference of the ideal binning and the maximally mixed state. The average is taken over typical circuits $\hat{U}$. ${\bf bins}$ is the array of summed probabilities computed by algorithms 1 and 2. We compute the weights of ${\bf bins}(\psi_{\rm ideal})$ with the pre-computed ideal probabilities $p_i$ and ${\bf bins}( {\mathbb I}/2^n)$ with the maximally mixed state ${\mathbb I}/2^n$. This method is sensitive to both incoherent and coherent errors.

We now extend the binning procedure to only detect incoherent errors. Instead of pre-computing the quantum state and binning each experimental output according to the ideal outcome, we bin the experimental output according to the experimentally measured frequencies (algorithm \ref{bog_algo_exp}).  The intuition behind this binning strategy is that since the Haar measure is invariant under unitary operations, the resulting measurement after a purely unitary error is still described by a Porter-Thomas distribution, just not the specific Porter-Thomas distribution as derived from the pre-computed quantum state. For this binning strategy, since the measured outcomes are not pre-assigned a bin, the weights of ${\bf bins}(\psi_{\rm ideal})$ are computed by integrating the ideal Porter-Thomas distribution and taking the integration bounds to be the edges of each bin, i.e. $\int_{a}^{b}q e^{-Nq}$. The weights of ${\bf bins}( {\mathbb I}/2^n)$ are computed similarly with the exception of integrating over a normalized Gaussian centered about $1/2^n$ whose width is proportional to $1/\sqrt{2^n\, \rm shots}$, i.e. $\int_{a}^{b} q\,{\rm exp}[\frac{-1}{2}(\frac{q-1/2^n}{\sigma})^2]$, where $\sigma=1/\sqrt{2^n\, \rm shots}$. A Gaussian is used here to account for shot noise, which is ubiquitous given a finite number of shots.


A simple example to give intuition for this method is to apply a bit flip error to each qubit at the end of the quantum circuit. The overall shape of the distribution is still a Porter-Thomas, but the labels of the bitstrings have been permuted. Binning these measurements according to the pre-computed quantum state, therefore, would yield a low fidelity. Binning by the experimental probabilities however, still yields a high fidelity as this methodology is blind to coherent errors.

We execute all of our experiments on a six qubit linear chain on $ibmq\_dublin$ (Fig. \ref{fig:2q_bog} (a)), a 27 qubit heavy-hex superconducting qubit processor. We first compare the results of two-qubit pairwise BOG and compare the CNOT error rate extracted in this way to the CNOT error rates measured by simultaneous RB \cite{Gambetta_2012}. These qubits are fixed frequency transmon with fixed nearest neighbor coupling where the native entangling gate is a CNOT mediated by an echoed cross-resonance microwave (CR) interaction \cite{magesan2020,sheldon2016_hamtomo}. We execute the circuit shown in Fig. \ref{fig:2q_bog} (b) among adjacent pairs in the chain, executing pair (Q2, Q3) and (Q5, Q8) simultaneously in one experiment and (Q1,Q2), (Q3, Q5), and (Q8, Q11) simultaneously in another experiment. By measuring adjacent qubit pairs simultaneously, cross-talk errors from applying simultaneous CR tones as well as spectator errors can be detected \cite{Gambetta_2012,sundaresan2020}. 

The experiments were repeated with 90 different seeds to generate 90 distinct circuits with unique random single-qubit rotations and each circuit was executed with 1000 shots. The shots from each experimental outcome were binned into 10 bins and summed over all seeds. As a rule of thumb, the number of seeds times the number of basis states in the Hilbert space should be much larger than the number of bins to build good statistics. The depths of the circuits were varied up to 270 cycles and the resulting fidelity curve was fit to an exponential decay in order to extract the error per cycle. We assume a two-qubit depolarizing noise channel such that the exponential decay rate is equal to the depolarizing parameter $\lambda$ and extract an error per gate equal to $\frac{3}{4}\lambda$, where the prefactor is due to the dimensionality of a two-qubit depolarizing channel \cite{Magesan11}. Fig. \ref{fig:2q_bog} (c)-(g) illustrates the CNOT error rates extracted by the BOG compared to the CNOT error rates measured by simultaneous RB the same day. We measure an average CNOT error rate of 1.292 (0.054)\% per gate from the BOG in comparison to 1.178 (0.025)\% per gate from RB, demonstrating agreement within 10\% between the two techniques. 

\begin{table}[t]
\begin{tabular}{ |p{1.5cm}||p{1.5cm}|p{1.5cm}|p{1.5cm}|p{1.5cm}|}
 \hline
 \hline
  & 2Q RB & 2Q BOG & \raggedright 2Q BOG (barriers) & 6Q BOG\\
 \hline
 \raggedright Avg. 2Q error rate & 1.178 (0.025)\% & 1.292 (0.054)\% & 1.340 (0.054)\% & 1.502 (0.022)\% \\
 \hline
 \hline
 \end{tabular}
 \caption{Average gate errors comparing between different benchmarking methods. Two-qubit BOG agrees within 10\% of two-qubit RB. Adding barriers between cycles in the two-qubit bog circuits enforces the CNOTs between different qubits to start simultaneously each cycle, resulting in idling qubits since CNOTs between different qubits have variable lengths. The resulting idle time increases the average gate error. The six-qubit BOG detects a higher average effective error rate due to unavoidable idle times during the alternating cycles of CNOTs.}
 \label{tab:error_rates}
 \end{table}

\begin{figure}
    \centering
    \includegraphics[width=\columnwidth]{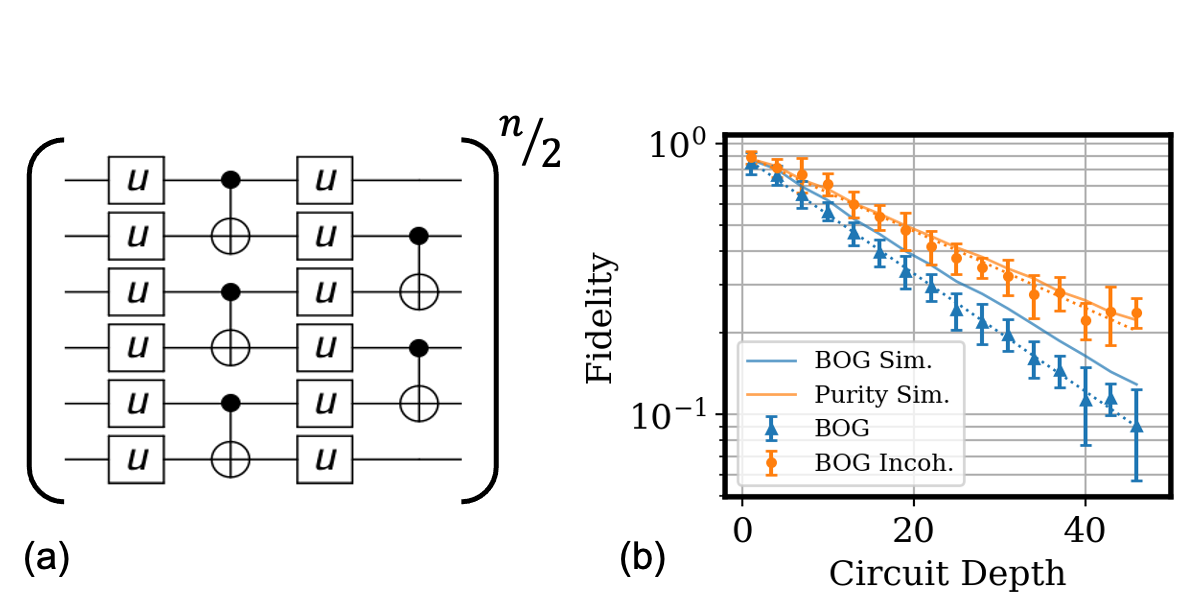}
    \caption{(a) Six-qubit BOG circuits. Haar random single-qubit rotations are applied to each qubit followed by CNOTs. Alternating layers of CNOTs are applied after single qubit rotations. (b) Data for the six-qubit BOG circuits. Data are binned according to quantum state (blue triangles) and are fit to an exponential to extract an average CNOT rate of 1.502 (0.022)\%. The blue solid line is a noisy simulation of the circuit with the CNOT errors measured from the two-qubit BOG circuits. The orange circles are the data binned by the experimental probability which detects only incoherent errors. Fitting these data to an exponential yields an average incoherent error rate of 0.980 (0.038) \%. The solid orange line is noisy simulation using error rates measured from purity RB.}
    \label{fig:6q_bog}
\end{figure}

Having established agreement between the RB and BOG in the two-qubit case, we now consider a six-qubit BOG experiment. In this regime, RB is difficult to perform because the decomposition of multi-qubit Clifford gates results in many more native entangling gates. Moreover, experiments beyond the two-qubit regime have been shown to capture error mechanisms absent from single and two-qubit measurements \cite{McKay19,mckay2020correlated}. 

The structure of the circuits executed are illustrated in Fig. \ref{fig:6q_bog} (a). Random single qubit rotations are applied to each qubit before applying CNOTs in parallel to adjacent qubit pairs, alternating each cycle which qubit pairs perform the CNOT, thus exploiting the native connectivity of our device via a hardware-efficient circuit structure. In general, any native entangling 2-qubit gate can be used in these circuits. These experiments are averaged over 40 random number seeds, repeated with 8000 shots each and the results were binned into 30 bins. Error bars are generated by calculating the standard deviation of eight groups of five seeds averaged together. Figure \ref{fig:6q_bog} (b) illustrates the fidelity of the six-qubit system as a function of the circuit depth, defined as one cycle of single qubit rotations and one cycle of parallel CNOTs. The blue points are the BOG fidelity binned by the pre-computed quantum state (algorithm \ref{bog_algo_sv}), which captures both incoherent and coherent noise. By fitting this curve to an exponential, we can relate the decay rate of the exponential, $\lambda$, to an average gate error ($EPG$) by $EPG=\frac{3}{4}\frac{2}{5}\lambda$. Here again the prefactor of $\frac{3}{4}$ is due to the dimensionality of a two-qubit depolarizing channel. The second prefactor $\frac{2}{5}$ is to account for the five CNOT gates per two circuit layers. In this six-qubit circuit we extract an average CNOT error rate of 1.502 (0.022)\%, notably higher than the average CNOT error rates measured in the two-qubit case, 1.292 (0.054)\%. These error rates are summarized in table \ref{tab:error_rates}. 

\begin{figure}[t]
    \centering
    \includegraphics[width=\columnwidth]{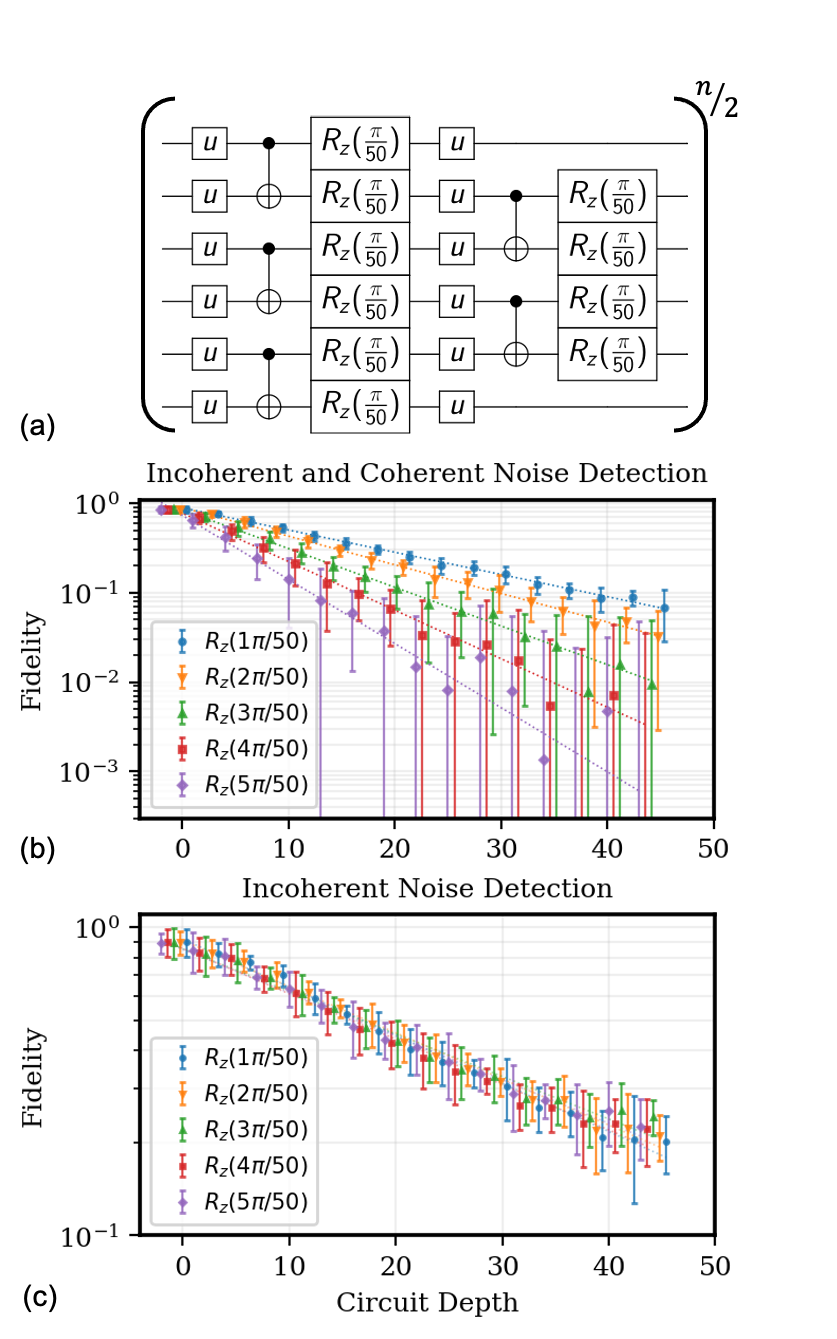}
    \caption{(a) Six-qubit BOG circuits with injected $Z$ noise following each CNOT. Virtual rotations of 1, 2, 3, 4, and 5\% of 2$\pi$ radians are injected as a purely coherent noise source. (b) Fidelity decay rates of the $Z$ noise injected circuits, binned by the precomputed quantum state (algorithm \ref{bog_algo_sv}), demonstrating a spread in decay rates due to the increase in coherent noise. Markers are offset in the x direction for clarity. (c) Fidelity decay rates of the same coherent noise injected circuits, binned by the experimental probability (algorithm \ref{bog_algo_exp}). Binning by this method only detects incoherent noise, and as such the decay rates of the five curves are identical. Markers have been offset in the x direction for clarity.}
    \label{fig:6q_bog_z}
\end{figure}

This increase in average error rate is likely due to the additional qubit idling time when the six-qubit BOG circuits are executed in comparison to the six-qubit BOG circuits. Idling qubits are introduced in two separate instances in the six-qubit BOG circuits. First, because the CNOT pairs now alternate between adjacent qubit pairs (Fig. \ref{fig:6q_bog} (a)) and because the CNOT lengths are different, the qubit pairs with the fastest CNOTs must wait for the slower CNOTs to finish executing before the next layer of the circuit can be executed. This is in contrast to the two-qubit BOG circuits (Fig. \ref{fig:2q_bog} (b)) where the start of the execution of a CNOT between qubits is not constrained to the timing of the adjacent CNOTs. Indeed, repeating the two-qubit BOG experiments with barriers inserted between the CNOT layers, which enforces the simultaneous timing of adjacent CNOTs, supports this conclusion resulting in a higher average CNOT error rate of 1.340 (0.054)\% (not shown). 

A second qubit idle period occurs when Q1 and Q11 are left idle during the even cycles of the six-qubit BOG circuits. This another error which is not captured by two-qubit measurements. On these cycles, CNOTs are only applied to the inner two pairs of qubits, (Q2, Q3) and (Q5, Q8). This highlights the ability to detect errors that are likely to occur during the execution of a realistic algorithm, but are not evident by one and two-qubit benchmarking.

\begin{figure}[t]
    \centering
    \includegraphics[width=0.9\columnwidth]{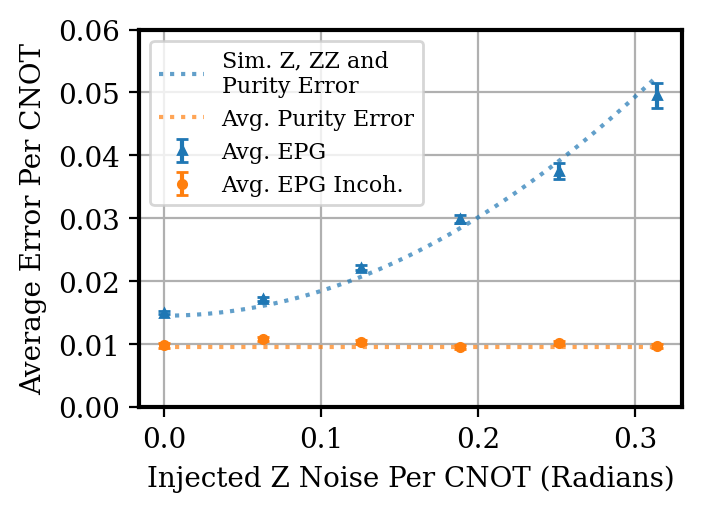}
    \caption{Extracted error average CNOT error rates from the six-qubit BOG experiments with injected coherent $Z$ noise. As the magnitude of the injected $Z$ noise increases, the average CNOT error increases quadratically, as the average incoherent CNOT error remains constant. Simulation of the gate (blue dotted line) with the injected $Z$ noise with an additional best fit $ZZ$ noise of 56.7 kHz is also shown. Additionally, the average purity error per gate is shown (orange dotted line). The difference in the curves indicates the amount of residual coherent noise.}
    \label{fig:z_noise_epg}
\end{figure}

We apply our second binning method to this six-qubit experiment (Fig. \ref{fig:6q_bog} (b)) in order to extract the average incoherent error per gate and compare this figure to the average two-qubit purity error as measured by purity RB \cite{wallman2015}. We demonstrate good agreement between the average incoherent error per gate as measured by the six-qubit BOG experiment, 0.980 (0.038)\%, compared with the average purity error per gate, 0.958 (0.023)\%. We note to build good statistics for this method, the number of shots taken must be much larger than the number of basis states in the Hilbert space. 


Finally, we repeat the six-qubit experiment while injecting a purely coherent $Z$ noise to each qubit each time a CNOT is applied to demonstrate the BOG is able to effectively discriminate between coherent and incoherent noise. After each CNOT, a small $Z$ rotation from 1\% to 5\% of a 2$\pi$ rotation is applied to the control and target qubits as a virtual rotation (Fig. \ref{fig:6q_bog_z} (a)). This injected noise takes 0 time as it is implemented in software as a reorientation of the x-axis of the Bloch sphere and can be executed with near perfect fidelity \cite{mckay2017_zgates}.

When binning by the pre-computed quantum state (Algorithm \ref{bog_algo_sv}), we observe an obvious increase in the decay rate of the fidelity curve which increases with the magnitude of the injected coherent noise (Fig. \ref{fig:6q_bog_z} (b)), as this method is sensitive to both incoherent and coherent errors. With increasing levels of coherent noise we observe a wider spread in the error bars. This effect is also commonly observed in RB when high levels of coherent noise are present because in certain circuits coherent noise can constructively or destructively interfere with itself, resulting in higher and lower fidelity outcomes. When binning by the experimentally measured probabilities (Algorithm \ref{bog_algo_exp}), the resulting decay rates remain constant as the injected noise is purely coherent and binning by this method only detects incoherent errors (Fig. \ref{fig:6q_bog_z} (b)).

The extracted error rates from Fig. \ref{fig:6q_bog_z} are summarized in Fig. \ref{fig:z_noise_epg}. In addition to the extracted average error per gate and incoherent error per gate, we plot the average purity error, measured from purity RB, and simulations of the average error per gate with coherent noise. As expected, the error per gate increases quadratically as the magnitude of the $Z$ noise increases, a hallmark of coherent noise \cite{sheldon2016_iterative,wallman2016}. The difference of these two curves indicates the relative amount of coherent noise in the gate error. Notably at 0 injected Z noise, there is a discrepancy between the average gate error and the incoherent gate error, indicating some amount of coherent noise that is unaccounted for. We model this residual noise as a static $ZZ$ interaction between the control and target qubit, applied after each CNOT. Fitting this $ZZ$ interaction with the average CNOT gate time of 443.73 ns, we extract an effective $ZZ$ interaction strength of 56.7 kHz, the same order of magnitude as the average independently-measured static $ZZ$ interaction of 24.4 kHz. With this simple model we capture the scaling behavior of the experimental data, though this effective $ZZ$ strength is likely an overestimate since it only considers the $ZZ$ interaction between the control and target qubits. The model omits the additional $ZZ$ interactions from nearest-neighbor spectator qubits and other higher weight interactions \cite{mckay2020correlated}, as well as the mitigating effects of echo pulses applied during the CNOT \cite{sundaresan2020}. Therefore, the $ZZ$ interaction we extract is likely some sum of all of these effects and our noise model is the simplest non-local model which reasonably reflects the experimental data.

In conclusion, we have demonstrated a hardware-efficient multi-qubit metric capable of discriminating between coherent and incoherent noise, a regime inaccessible to current benchmarking techniques. This technique can be used as a diagnostic tool for different hardware platforms as the BOG accommodates different native two-qubit entangling gates and requires no special qubit connectivity or layout. It can determine whether the algorithmic performance of a quantum computing system is limited by its coherence or by the quality of the control and calibrations. If the system fidelity is limited by unitary errors, it can determine how much of the error can potentially be corrected by improvements in control. In addition, it can detect error sources that are only evident in multi-qubit experiments (namely true quantum algorithms), errors that may not be evident from standard two-qubit benchmarks. As we move towards circuits for error correcting codes, the ability to detect coherent vs. incoherent errors is critical. There are also further open questions about directly connecting the metrics in this manuscript, or any metrics derived from random circuits, to the performance of large scale codes. A possible future extension is to replace the random unitary operators in our circuit construction with random elements of the Clifford group, producing circuits that are analogous to those used in stabilizer measurements.


This work was sponsored by the Army Research Office
and was accomplished under
Grant Numbers W911NF-14-1-0124 and W911NF-21-1-
0002. The views and conclusions contained in this document are those of the authors and should not be interpreted as representing the official policies, either expressed or implied, of the Army Research Office or the
U.S. Government. The U.S. Government is authorized
to reproduce and distribute reprints for Government purposes notwithstanding any copyright notation herein



\bibliography{mainbib}

\end{document}